\documentclass{svmult}
\usepackage{graphicx}

\newcommand{\ba}{\begin{array}}
\newcommand{\ea}{\end{array}}
\newcommand{\bd}{\begin{displaymath}}
\newcommand{\ed}{\end{displaymath}}
\newcommand{\be}{\begin{equation}}
\newcommand{\ee}{\end{equation}}
\newcommand{\bea}{\begin{eqnarray}}
\newcommand{\eea}{\end{eqnarray}}

\def\barr{\begin{array}}
\def\earr{\end{array}}

\newcommand{\beq}{\begin{eqnarray}}
\newcommand{\eqn}{\end{eqnarray}}

\def\non{\nonumber}

\def\etal{ {\em et al.}}

\def\q2 {q^2}

\def\N10{\widetilde \chi_1^0}
\def\Cp1{\widetilde \chi_1^+}
\def\Cm1{\widetilde \chi_1^-}
\def\C1pm{\widetilde \chi_1^\pm}

\begin{document}
\title*{New Source of CP violation in $B$ physics ?}
\author { N.G. Deshpande and Dilip Kumar Ghosh}
\institute{Institute of Theoretical Science \\
University of Oregon, Eugene, OR 97403}

\maketitle
In this talk we discuss how the down type 
left-right squark mixing in Supersymmetry
can induce a new source of CP violation in the time dependent 
asymmtries in $B \to \phi K$ process. We use QCD improved factorization
process to calculate the hadronic matrix element for the process and
find the allowed parameter space for $\rho $ and $\phi $, the magnitude 
and phase of the  down type LR(RL) squark mixing parameter 
$\delta^{bs}_{LR(RL)}$. In the same allowed regin we calculate the
expected CP asymmtries in the $B \to \phi K^{*}$ process. 
 



\section{Introduction}

Time dependent asymmetries measured in the decay $B \to \phi K_S$ both by 
BaBar and Belle collaborations \cite{babar, belle1, belle2,hamel} show 
significant deviation 
from the standard model and this has generated much theoretical speculation 
regarding physics beyond the standard model [\ref{np1}-\ref{huitu}]. 
In the standard
model, the process $B \to \phi K_S$ is purely penguin dominated and the
leading contribution has no weak phase. The coefficient of 
$\sin(\Delta m_B t) $ in the asymmetry therefore should measure 
$\sin 2\beta $, the same quantity that is involved in $B \to \psi K_S$ 
in the standard model. The most recent measured average values of 
asymmetries are ~\cite{hamel,browder}
\beq
S_{\psi K_S} & = & 0.734 \pm 0.055\non\\
S_{\phi K_S} & = & -0.15 \pm 0.33 
\label{sin2beta}
\eqn
The value for $ S_{\psi K_S} $ agrees with theoretical expectation from the
CKM matrix of $S_{\psi K_S} = \sin 2\beta = 0.715^{+0.05}_{-0.045}$
~\cite{buras1}. This leads to the conclusion that CP phase in $B-\bar B$ 
mixing is consistent with the standard model. 
The deviation in the $\phi K_S$ is intriguing because a penguin process 
being a 
loop induced process is particularly sensitive to new physics which can 
manifest itself in a loop diagram through exchange of heavy particles. 
In this talk \cite{self} we consider effects arising from non universal squark
mixing in the second and third generation of the down type squarks 
in supersymmetric theory as the 
origin of additional contributions to the amplitude within the mass insertion
approximation scheme. 
In particular, the exchanges of gluinos $(\tilde g)$ and squark $(\tilde q)$
with left-right mixing can enhance the Wilson coefficient of the 
gluonic dipole penguin operator ${\cal O}_{8g}$ by a factor of 
$m_{\tilde g}/m_b$ compared with the standard model prediction 
and we take into account its effect on the process $B \to \phi K_S$. 
In our analysis
we take the $B -\bar B$ mixing phase the 
same as in the standard model as required
by $\psi K_S$ data, and permitted in SUSY by requiring that the first and 
third generation squark mixing to be small.
We study $B \to \phi K$ in QCD improved factorization
scheme (BBNS approach ) \cite {beneke}. This method incorporates elements of
naive factorization approach (as its leading term ) and perturbative QCD 
corrections (as sub-leading contributions) and 
allows one to compute systematic
radiative corrections to the naive factorization for the hadronic $B$ decays.

In supersymmetry, assuming masses of squarks $(\tilde q)$ 
 and gluinos $(\tilde g )$,
the new source of CP violation can be parameterized by the complex quantity 
$\delta^{bs}_{LR(RL)}$ written in the form $\rho e^{i \psi}$
We identify
 the region in $\rho- \psi $ plane allowed by the experimental data
on $B \to \phi K$ time dependent asymmetries $S_{\phi K_S}$ and 
$C_{\phi K_S}$ and the branching ratio. 
This allowed region is dependent of the QCD scale $\mu$, therefore we 
illustrate
the region for two values of $\mu = m_b $ and $ m_b/2 $. 
The same contribution should also be present in other penguin mediated process.
We study the effect of $LR(RL)$ mass insertion to the 
$B \to \phi K^{\ast}$  decay mode which is also a pure penguin 
process using QCD improved 
factorization method.  We then estimate the branching ratio 
${\cal B}(B \to \phi K^\ast)$ and the CP asymmetry ${\cal A}_{CP}$ in the 
parameter space of $\delta^{bs}_{LR(RL)}$ allowed by $B\to \phi K$ data.
In this vector vector final state, one can also construct more
CP violating observables \cite{sinha-london}. We compute these observables in 
the same range of parameter space as that allowed by $B \to \phi K$.

\section{CP Asymmetry of $B \to \phi K $}

The time dependent CP asymmetry of $B \to \phi K_S$ is described by :
\beq
{\cal A}_{\phi K_S}(t) &=& \frac{\Gamma(\overline{ B^0}(t) \to \phi K_S) 
- \Gamma (B^0(t) \to \phi K_S)}{ \Gamma(\overline{ B^0}(t) \to \phi K_S)+
\Gamma (B^0(t) \to \phi K_S)}\\
&=& -C_{\phi K_S} \cos (\Delta m_B t ) + S_{\phi K} \sin(\Delta m_B t )
\eqn
where $S_{\phi K}$ and $C_{\phi K_S} $ are given by 

\beq
S_{\phi K} = \frac{2 Im~\lambda_{\phi K_S}}{1 + \mid \lambda_{\phi K_S}\mid^2}
,~~~~ 
C_{\phi K_S} = \frac{1- \mid \lambda_{\phi K_S}\mid^2}
{1 + \mid \lambda_{\phi K_S}\mid^2}
\eqn
and $\lambda_{\phi K_S}$ can be expressed in terms of decay amplitudes:
\beq
\lambda_{\phi K_S} = -e^{-2i \beta }\frac{{\overline {\cal M}}(\overline{B^0 }\to \phi K_S)}{{\cal M}(B^0 \to \phi K_S)}
\eqn

The branching ratio and the direct CP asymmetries of both the charged and 
neutral modes of $B \to \phi K$ have been measured 
\cite{babar, belle1, belle2,hamel,browder,CLEO}:
\beq
{\cal B}(B^0 \to \phi K_S) & =& (8.0\pm 1.3) \times 10^{-6}\\
{\cal B}(B^+ \to \phi K^+) & =& (9.4 \pm 0.9) \times 10^{-6} ,\\
S_{\phi K_S} & = & +0.45\pm 0.43 \pm 0.07~~(\rm {BaBar}); \\ 
             & = & -0.96\pm 0.50^{+0.09}_{-0.11}~~({\rm Belle });\\
C_{\phi K_S} & = & -0.19 \pm 0.30 \\
{\cal A}_{CP}(B^+ \to \phi K^+) & = & (3.9 \pm 8.8 \pm 1.1)\% 
\eqn

\section{The exclusive $B\to \phi K $ decay }

\begin{figure}[t!]
\begin{center}
\setlength{\unitlength}{1cm}
\includegraphics[clip,scale=1]{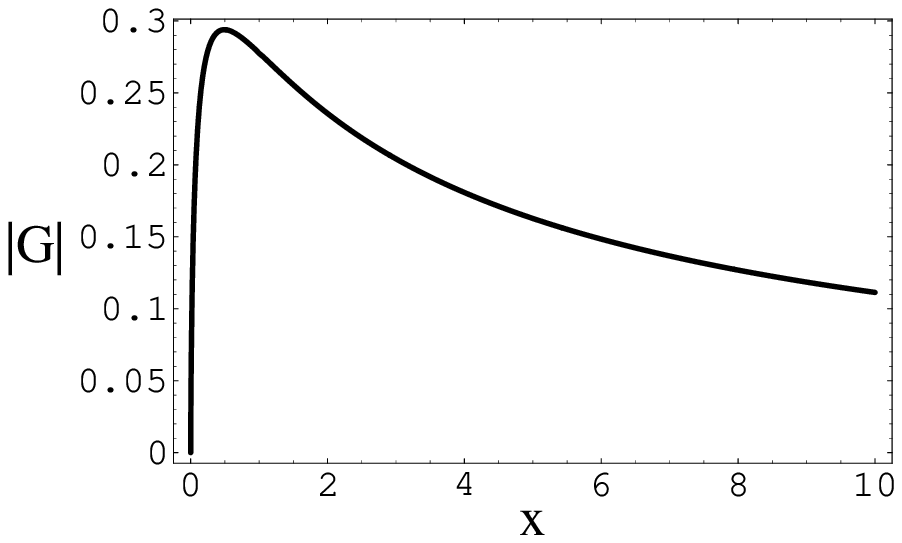}
\end{center}
\vspace*{-12.0cm}
\caption{Variation of $\mid G(x)\mid$ with $x(= m^2_{\tilde g}/m^2_{\tilde q})$.}
\label{fig0}
\end{figure}
In the standard model, the effective Hamiltonian for charmless 
$B \to \phi K (\phi K^\ast) $ decay is given by \cite{beneke} 
\beq
{\cal H}_{eff} &=& -\frac{G_F}{\sqrt{2}} V_{tb}V^{\ast}_{ts}
\Bigg \{ C_1(\mu) {\cal O}_1(\mu) + C_2(\mu) {\cal O}_2(\mu) 
+ \sum^{10}_{i= 3} C_i(\mu) {\cal O}_i(\mu)  + C_{7 \gamma} {\cal O}_{7 \gamma}
\nonumber \\ 
&&
+ C_{8 g} {\cal O}_{8 g} \Bigg \}
\eqn 
where the Wilson coefficients $C_i(\mu)$ are obtained from the weak scale 
down to scale $\mu$ by running the renormalization group equations. The 
definitions of the operators and different Wilson coefficients 
can be found in Ref.\cite{beneke}.

\section{ $B \to \phi K $ in the QCDF Approach}

In the QCD improved factorization scheme, the $B \to \phi K $ decay amplitude 
due to a particular operator can be represented in following form :
\beq
< \phi K \mid {\cal O}\mid B> = < \phi K \mid {\cal O}\mid B>_{fact}
\left [1 + \sum r_n \alpha_{s}^n + O(\Lambda_{QCD}/m_b)\right ] 
\eqn
where $< \phi K \mid {\cal O}\mid B>_{fact}$ denotes the naive factorization 
result. The second and third term in the bracket represent higher order 
$\alpha_s$ and $ \Lambda_{QCD}/m_b $ correction to the hadronic transition
amplitude. Following the scheme and notations presented in 
Ref.\cite{jpma, huang}, we write down the $B \to \phi K $ amplitude in the
heavy quark limit.
\beq
{\cal M }(B^+ \to \phi K^+ ) &=&
\nonumber \\  
{\cal M }(B^0 \to \phi K^0 )
&=&\frac{G_F}{\sqrt{2}}m_B^2 f_{\phi} F_1^{B \to K}(m^2_{\phi})
V_{pb}V^{\ast}_{ps}\left [ a^p_3 + a^p_4+ a^p_5 
\right.
\nonumber \\ 
&&
\left.
-\frac{(a^p_7+a^p_9+a^p_{10})}{2} + a^p_{10a} \right ]
\eqn
where $p$ is summed over $u$ and $c$. The coefficients $a^p_i$ are given by 
\begin{eqnarray}
&&a^u_3 = a^c_3 =  C_3 + {C_4 \over N_c}
\left[ 1 + {C_F \alpha_s \over 4 \pi} \left (V_{\phi} + H_{\phi}\right )\right],
\nonumber\\ 
&&a^p_4  = 
C_4 + {C_3\over N_c} \left[ 1 + {C_F \alpha_s \over 4 \pi} 
\left (V_{\phi} + H_{\phi}\right ) \right] + 
{C_F \alpha_s \over 4\pi N_c} P^p_{\phi}, 
\nonumber\\ 
&&a^u_5  = a^c_5=  C_5 + {C_6 \over N_c}
\left[ 1 + {C_F \alpha_s \over 4 \pi} 
(-12-V_{\phi}) \right],
\nonumber\\ 
&&a^u_7  =  a^c_7 = C_7 + {C_8 \over N_c}
\left[ 1 + {C_F \alpha_s \over 4 \pi} 
(-12-V_{\phi} - H_{\phi}) \right],
\nonumber\\ 
&&a^u_9  = a^c_9 =  C_9 + {C_{10} \over N_c}
\left[ 1 + {C_F \alpha_s \over 4 \pi} 
\left (V_{\phi} + H_{\phi}\right ) \right],
\nonumber\\ 
&&a^u_{10}  =  a^c_{10} = 
\left[ 1 + {C_F \alpha_s \over 4 \pi} \left (V_{\phi} + H_{\phi}\right )
\right],
\nonumber \\
&& a^u_{10a} = a^c_{10a}= {C_F \alpha_s \over 4 \pi  N_c} Q_{\phi}
\end{eqnarray}
with $C_F = (N^2_c-1)/2N_c$ and $N_c = 3$.
The quantities $V_{\phi}, H_{\phi}, P^p_{\phi}$ and $Q^p_{\phi}$ are hadronic
parameters that contain all nonperturbative dynamics.
\beq
V_{\phi} &=& -12 \ln{\frac{\mu}{m_b}}- 18  + f^I_{\phi},
\nonumber \\
f^I_{\phi} &=& \int^1_0 dx g(x) \Phi_{\phi}(x);~~~~~ 
g(x) = 3 \frac{1-2 x}{1-x}\ln{x} - 3 i \pi,
\nonumber\\
H_{\phi} & = &\frac{4 \pi^2}{N_c} \frac{f_B f_{K}}{
F_1^{B \to K}(0) m^2_B} 
\int^1_0 dz \frac{\Phi_{B}(z)}{z}
\int^1_0 dx \frac{\Phi_{K}(x)}{x}
\int^1_0 dy \frac{\Phi_{\phi}(y)}{y},
\nonumber \\
P^p_{\phi} &=& C_3 \left [G_{\phi}(s_s)+ G_{\phi}(s_b) \right ] 
+ C_2 G_{\phi}(s_p) + (C_4+C_6)\sum_{f= u}^b {\tilde G}_{\phi}(s_f) 
+ C^{eff}_{8g} G_{\phi g}
\nonumber \\
Q_{\phi} & = & (C_8+C_{10})\frac{3}{2}\sum_{f=u}^b e_f G_{\phi}(s_f)
+C_9 \frac{3}{2} \left[ e_s G_{\phi}(s_s) + e_b G_{\phi}(s_b)\right ]
\nonumber \\
G_{\phi}(s) &=& \frac{2}{3} -\frac{4}{3}\ln{\frac{\mu}{m_b}} +
4 \int^1_0 dx~\Phi_{\phi}(x) \int^1_0 du~u~(1-u)\ln\left[s - u(1-u)(1-x)\right]
\nonumber \\
{\tilde G}_{\phi}(s) & = & G_{\phi}(s) - (2/3) \nonumber \\
G_{\phi g} & = & -\int^1_0 dx \frac{2}{(1-x)}\Phi_{\phi}(x)
\label{bphik}
\eqn
where, $s_i = m_i^2/m_b^2 $.
Here, $V_{\phi}$ represent contributions from 
the vertex correction and $H_{\phi}$ correspond to hard 
gluon-exchange interactions with spectator quarks. 
$P^p_{\phi}$ and $Q^p_{\phi}$ represent QCD penguin contributions.
We neglect order $\alpha_{em}$ EW penguin corrections to $a_i$.
$f_B, f_K $ are the $B$ and $K$ meson decay constants and $F^{B\to K}_{1}$
denotes the form factor for $B \to K$ transitions. 
$\Phi_B(z), \Phi_{K}(x) $, and $\Phi_{\phi}(y)$ are the $B, K $, and $\phi$
meson wave functions respectively. In this analysis we take following forms
for them \cite{jpma}
\beq
\Phi_B(x) &=& N_B~x^2~(1 - x)^2 \exp\left[-\frac{m^2_B x^2}{2 \omega^2_B}\right],
\nonumber\\
\Phi_{K,\phi}(x) &= & 6~x~( 1- x) 
\eqn
where, $N_B$ is a normalization factor satisfying $\int^1_0 dx~\Phi_B(x) =1 $,
and $\omega_B = 0.4 $.

For the sake of completeness, we give the branching ratio for $B\to \phi K$ 
decay channel in the rest frame of the $B$ meson.
\beq
{\cal BR}(B\to \phi K) = \frac{\tau_B}{8 \pi}\frac{\mid P_{cm} \mid}{m^2_B}
\mid {\cal M}(B \to \phi K) \mid^2  
\eqn
where, $\tau_B$ represents the $B$ meson lifetime and the kinematical factor
$\mid P_{cm}\mid $ is written as 
\beq
\mid P_{cm} \mid = \frac{1}{2 m_B}\sqrt{\left [m^2_B - (m_K+m_\phi)^2\right]
\left [m^2_B -(m_K - m_\phi)^2\right] } 
\eqn

\section{SUSY gluino contributions to $B \to \phi K $ }
 
In order to study the new physics contribution to the CP violating
phase of amplitude ${\cal M}(B \to \phi K)$, we compute the effect of 
flavor changing contribution
to $B \to \phi K$ arising from $q -\tilde q - \tilde g $ interactions in
supersymmetric theory under the mass insertion approximation scheme 
\cite{mass1, mass2}. In this approximation, the flavor changing contribution
is parameterized in terms of 
$\delta^{ij}_{AB} = \Delta^{ij}_{AB}/{\tilde m}^2 $, where, $\Delta$ 
represents the off-diagonal entries of the squark mass matrices, ${\tilde m}$
is an average squark mass, $A,B = L, R$ and $i,j$ are the generation 
indices. The $LR(RL)$ mass insertion can enhance the 
Wilson coefficients $C_{7\gamma}$ and $C_{8g}$ by a factor 
of $m_{\tilde g}/m_b$ compared to the standard model
contribution. This leads to a strong limit of order $O(10^{-2})$ on the 
$LR (RL) $ insertions $\mid \delta^{bs}_{LR(RL)}\mid$
from the ${\cal B}(B \to X_s\gamma)$ 
\cite{mass2,bsg1} while the limit on the $LL$ and $RR$ ones is
rather mild \cite{mass2,bsg1}. Thus, although larger values for $LL$ and 
$RR$ mixings are allowed, when one considers $B \to \phi K$, the effect of 
their mixings are only significant in the parameter space where the 
squark and gluino masses are at the edge of their experimental constraints
\cite{glkane}. Motivated by this fact, we only 
concentrate on $LR(RL)$ down type squark mixing in hereafter. 
Thus, the new physics effect is very sensitive to 
$\delta^{bs}_{LR(RL)}$.

In general, these contributions $LR(RL)$ can generate gluonic dipole 
interactions with the same as well as opposite chiral structure as the 
standard model. In our analysis we will consider 
each of them separately. Furthermore we will only consider the 
gluonic dipole moment operator, which is the dominant operator for this 
process.    
 
The effective Wilson coefficient for $C^{SUSY}_{8 g}$ obtained in the mass
insertion approximation is given by for the same chiral structure as the
standard model \cite {BCIRS,HE1}
\begin{eqnarray}
C_{8g}^{SUSY} (m_{\tilde q}) & = & - \, \frac{\sqrt{2} \pi
\alpha_s}{G_F (V_{ub} V_{us}^* +
V_{cb} V_{cs}^*) m_{\tilde{g}}^2} \, \delta^{bs}_{LR(RL)} \,
\frac{m_{\tilde{g}}}{m_b} \, G_(x) \, ,
\end{eqnarray}
with
\begin{eqnarray}
G(x) & = &
\frac{x}{3(1-x)^4} \,
\left[ 22-20x-2x^2+16 x \ln(x) -x^2 \ln (x)
+ 9 \ln (x) \right] ,
\end{eqnarray}
where $x=m_{\tilde{g}}^2 / m_{\tilde{q}}^2$ is
the ratio of the gluino and squark mass.

Using the renormalization group equation one can evolve the 
coefficient $C^{SUSY}_{8g}$ from the high scale $m_{\tilde q}$ to the
scale $m_b$ relevant for $B \to \phi K $ decay \cite{BCIRS}
\begin{eqnarray}
C_{8g}^{SUSY} (m_b) & = &
\eta C_{8g}^{SUSY}(m_{\tilde{q}} ) \, ,
\end{eqnarray}
with
\begin{equation}
\eta \, = \,
\left( \alpha_s(m_{\tilde{q}})/
\alpha_s(m_t) \right)^{2/21}
\left( \alpha_s(m_t)/
\alpha_s(m_b) \right)^{2/23}
\end{equation}
One can obtain $C^{SUSY}_{8g}$ for opposite chirality, by adding one more 
operator similar to ${\cal O}_{8g}$ with $(1+ \gamma_5) \to (1-\gamma_5)$ and
$\delta^{bs}_{LR} \to \delta^{bs}_{RL}$. However, in $B \to \phi K $ process,
both $LR$ and $RL$ contribute with the same sign because $B$ and $K$ 
parity are both $0^{-}$, and the process is parity conserving. 

The effective 
Wilson coefficient $C^{eff}_{8g}$ is defined as 
$C^{eff}_{8g} = C_{8g} + C^{SUSY}_{8g}$. 
This effective $C^{eff}_{8g}$ will contribute
to the amplitude ${\cal M}(B \to \phi K)$ through the function $P^p_{\phi}$ of
Equation \ref{bphik}. $C^{eff}_{8g}$ depends on the magnitude and phase of
the $(\delta^{bs}_{LR(RL)})$, 
value of squark mass $(m_{\tilde q})$ and the ratio 
$x~( = m^2_{\tilde g}/m^2_{\tilde q})$. The variation of $C^{eff}_{8g}$ with
$x$ is determined by the function $\mid G(x) \mid 
$ as shown in Figure~\ref{fig0}. From 
this Figure, it is clear that SUSY gluino contribution to $B \to \phi K $
first increases with increase in $x$, and then after some value of $x= 0.5$, 
it starts decreasing asymptotically with further increase in $x$. 

The different input parameters and their values used in numerical calculation 
of branching ratio and CP asymmetries are given in Ref.\cite{self}. 

 \subsection {$LR(RL) $ mixing}
In this section we study the effect of $LR(RL)$ mixing in 
$B \to \phi K$ process.
This $LR(RL)$ mixing of the down type squark sector 
can also affect the $ B \to \gamma X_s$ process and $B_s -\bar B_s$ mixing.
Hence we need to take into account the limit on $LR(RL)$ mixing parameter 
$\delta^{bs}_{LR(RL)}$ from the above two experimental data  
in the present analysis. In the first case, it has been shown 
in Ref.~\cite{mass2} that from 
the measurement of ${\cal B}( B \to \gamma X_s)$ one gets 
$\mid \delta^{bs}_{LR(RL)}\mid  < 
1.0 \times 10^{-2}$ and $3.0 \times 10^{-2}$ for $x = 0.3 $ and 4 respectively,
with $m_{\tilde q} = 500 $ GeV. It is interesting to note that the lower the
$x$ value stronger the limit on $ \mid \delta^{bs}_{LR(RL)}\mid $, which can 
be explained by the $x$ dependent behavior of the $C^{SUSY}_{7\gamma}$. 

The current experimental data on $B_s-\bar B_s$ mixing is $\Delta M_s > 14.4 $ ${\rm ps}^{-1}$ ( at $95\%$ C.L.) \cite {stocchi}. We have found that the 
$LR(RL)$ mixing does not change the value of $\Delta M_s$ significantly 
from the standard model prediction in the allowed range of 
$\mid \delta^{bs}_{LR(RL)}\mid $. 

\begin{figure}[t!]
\begin{center}
\setlength{\unitlength}{1cm}
\includegraphics[clip,scale=1]{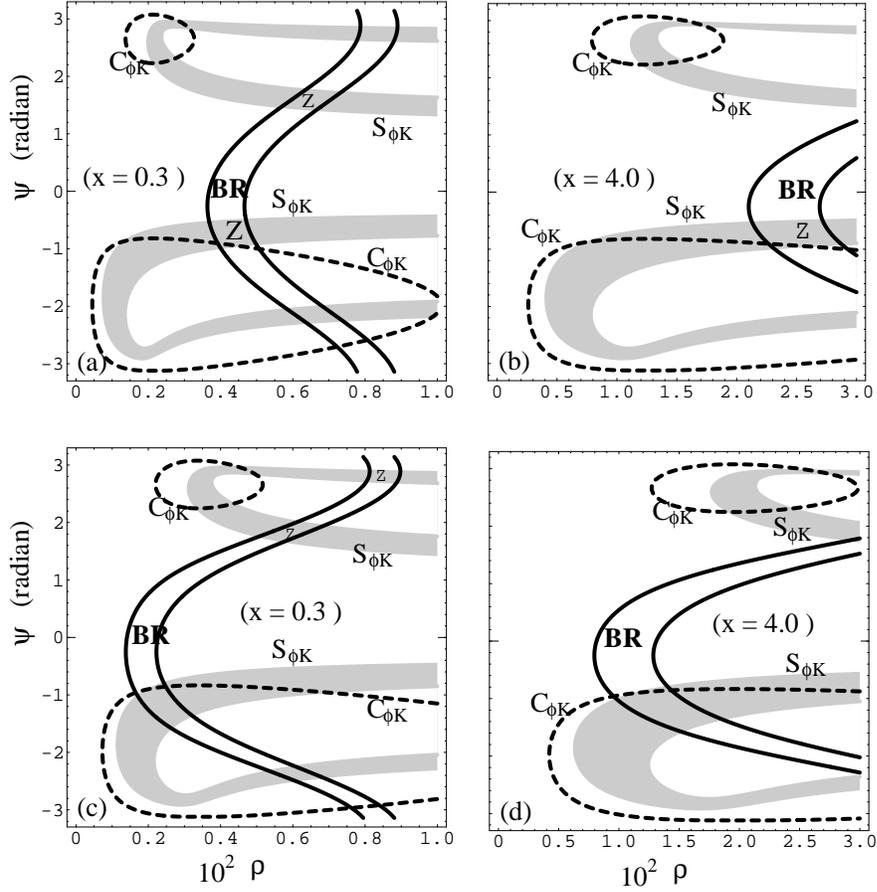}
\end{center}
\vspace*{-11.0cm}
\caption{Contour plots of $S_{\phi K}, C_{\phi K}$ and 
${\cal B}(B \to \phi K_S)$ in $\rho -\psi$ plane for two 
values of $x= 0.3~(a, c) $ and $4.0~(b, d)$ for $LR(RL)$ mixing with 
$m_{\tilde q} = 500 $ GeV. 
The scale $\mu = m_b$ for Figures $(a)$ and $(b)$, 
while it is $m_b/2$ for Figures $(c)$ and $(d)$. 
The $1\sigma $ allowed regions of $S_{\phi K}$, ${\cal B}$
and $C_{\phi K}$ are two gray bands, area within the solid curves and
area outside the two dotted contours respectively.}
\label{fig1}
\end{figure}

In our analysis we consider $m_{\tilde q} = 500 $ GeV and 
take two values of $x= 0.3 $ and $4.0$, which will determine the gluino
masses.  
In Figure \ref{fig1} we show the  
$1\sigma $ allowed region in $\rho-\psi $ plane from $B \to \phi K$ data on
$S_{\phi K}, C_{\phi K} $ and ${\cal B}$. 
The gray band indicate the parameter space which is allowed 
by $S_{\phi K}$. The area outside the two dotted contours 
is allowed by $C_{\phi K}$, while the area enclosed by the solid curves is 
allowed by the ${\cal B}( B\to \phi K_S)$ measurement.      
The region (marked by $Z$) in gray band enclosed by the solid curves 
is the only parameter space left in $\rho-\psi$ plane which is 
allowed by the experimentally measured $S_{\phi K}, C_{\phi K}$  
and ${\cal B}$ within $1\sigma $.

The Figures~\ref{fig1}$(a)$ and $(b)$ , correspond to contour plots for 
$x = 0.3 $ and $4.0$ respectively at the scale $\mu = m_b$. 
For $x =0.3$, we get two  
allowed regions each at positive and negative values of the 
new phase $\psi $. On
the other hand for $x = 4.0$, we get only one allowed region which lies 
at the negative value of $\psi$ and at much higher value of 
$\rho> 2.2\times 10^{-2} $. We have noticed before that the constraint on 
$LR(RL)$ mixing parameter from the ${\cal B}(B \to X_s \gamma )$ is stronger
at $x = 0.3$ compared to the limit at $x = 4.0$. This behavior is also 
reflected in the $B \to \phi K$ process, 
 where we find that, for $x = 4.0$,   
the $1\sigma $ constraint from $S_{\phi K}, C_{\phi K}$ and ${\cal B}$ 
is much weaker compared to the constraint shown in 
Figure~\ref{fig1}$(a)$ correspond to $x= 0.3 $. 

Similar allowed regions are shown in Figure \ref{fig1}$(c)$ for 
a different choice of the QCD scale $\mu = m_b/2$. One can see that 
the allowed parameter space does depend on $\mu$. In this case, both 
the allowed regions are confined at the positive value of $\psi$.  
For $x = 4.0$ ( Figure \ref{fig1}$(d)$), there are no allowed regions. From 
the $S_{\phi K}$ and branching ratio contour one can see that the 
allowed region from $B \to \phi K$ require some higher value of $\rho$ which
lies beyond $B\to X_s\gamma $ limit.  

Before we conclude this section, we would like to compare our predictions
with some of the existing literatures on $B \to \phi K$
process \cite{harnik, ciuchini2, glkane}.  

We agree qualitatively with the results of 
Ref.\cite{ciuchini2,glkane} in 
places where we overlap. Similar to our approach, both of these analyses 
were based upon the QCD improved factorization scheme. However, there are
some quantitative differences between these papers 
and our analysis. For example, we differ in the choice of squark and gluino 
masses, the authors of the above two papers considered degenerate squark and 
gluino masses, whereas we have considered non-degenerate squark-gluino masses.
We have fixed the squark mass at 500 GeV and considered two values of 
the gluino masses, 
determined by the parameter $x$ defined earlier. 
Secondly, we have performed our analysis for two values the QCD scale, 
$\mu = m_b/2$ and $m_b$. Our results depend strongly on the choice of the
ratio $x$ and also on the scale $\mu$. However, in a broad sense, we do agree 
that to satisfy $B\to \phi K $ data, one requires 
$\mid \delta^{bs}_{LR(RL)}\mid \sim 10^{-3}- 10^{-2}$. 

In Ref.\cite{harnik}, authors made a detailed investigation of a scenario 
in which the $LR$ and $RR$ operators co-exist. Moreover, because of the large
mixing, the calculation was done in the mass eigenbasis with more model
dependence than ours. It has been shown in this analysis that $RR$ 
insertion (which arises due to a large mixing between $\tilde s_R$ and 
$\tilde b_R$) could show sizable effect on $S_{\phi K}$, but only for very
light gluino mass, near the experimental bound. Such a large $RR$ mixing
also modify $\Delta M_{s}$ significantly which can be observed at the 
Tevatron Run II. In their second case, they have the 
combination of both large right-right and left-right squark mixing 
($LR + RR)$. In this case the squark and gluinos could be sufficiently 
heavy to have no significant enhancement of the $\Delta M_s $.     
  
From our analysis we observe that SUSY leads to a comprehensive 
understanding of $B \to \phi K_S$ data though in a very limited parameter 
space of $\delta^{bs}_{LR(RL)}$. In rest of the paper we now explore the 
consequence of such $LR(RL)$ mixing of squarks in the $B \to \phi K^\ast$ 
process.
 
\section {$B \to \phi K^{\ast}$ decay } 
In this section we will study the effect of $LR(RL)$ mixing of down type
squarks to $B \to  \phi K^{\ast} $ process through the gluonic 
dipole moment operator $C_{8g}$. 
We will study the $B \to \phi K^\ast$ 
process by using the QCD improved factorization. Using this method one
can compute nonfactorizable corrections to the above process in the
heavy quark limit. Recently the $B \to VV $ process has been computed 
using QCD improved factorization method \cite{cheng-yang1}. 
In rest of analysis we will follow Ref. \cite{cheng-yang1}. 
  
The most general Lorentz invariant decay amplitude for the 
process $B \to VV$ can be expressed as 
\beq
{\cal M}(B(p_B) \to V_1(\epsilon_1, p_{1}) V_2(\epsilon_2, p_2))\propto 
\epsilon^{\ast\mu}_1 \epsilon^{\ast\nu}_2 \left [ a g_{\mu\nu} 
+ b p_{B\mu}p_{B\nu} 
+ i c \epsilon_{\mu\nu\alpha\beta} p^\alpha_1 p^\beta_2 \right ]
\eqn
where the coefficients $c$ correspond to the $p$-wave amplitude, and $a,b$
to the mixture of $s$ and $d$ wave amplitudes. Using these $a,b$ and $c$ 
coefficients one can 
construct the three helicity amplitudes:
\beq
H_{00} &=& \frac{1}{2 m_{V_1} m_{V_2}}\left [ ( m^2_B - m^2_{V_1} - m^2_{V_2})
 a + 2 m^2_B p^2_{cm} b \right ]\ \nonumber\\
H_{\pm \pm} &=& a \mp m_B p_{cm} c
\eqn
where $p_{cm}$ is the center of mass momentum of the vector meson in the $B$
rest frame and $m_{V_1}(m_{V_2})$ is the mass of the vector meson $V_1(V_2)$.
These helicity amplitudes $H_{00}$ and $H_{\pm\pm}$ can be related to the spin
amplitudes in the transverse basis $(A_0, A_{\mid\mid}, A_{\perp})$ defined in
terms of linear polarization of the vector mesons:
\beq
A_{0} &=& H_{00}\non\\
A_{\mid\mid}&=& \frac{1}{\sqrt{2}}\left ( H_{++} + H_{--}\right ) \non\\
A_{\perp} &=& \frac{1}{\sqrt{2}}\left ( H_{++} - H_{--}\right ) 
\eqn
\begin{table}
\begin{center}
\begin{tabular}{|c|c|c|}
\hline
Branching ratio  &  Data & Weighted average \\
\hline
$B^+ \to \phi K^{*+}  $ &BaBar~~~~ 
$ \left (12.1^{+2.1}_{-1.9}\pm 1.5\right)\times 10^{-6} $ & \\
  &CLEO ~~~~ $ \left(10.6_{-4.9-1.6}^{+6.4+1.8}\right )\times 10^{-6} $ 
& $\left (9.9 \pm 1.23 \right)\times 10^{-6}$\\
           &Belle~~~~ $ \left (9.4\pm1.1\pm0.7\right)\times 10^{-6}$ & \\
\hline
$ \bar{B}^0 \to \phi \bar{K}^{*0} $ &BaBar~~~~ 
$ \left( 11.1_{-1.2}^{+1.3} \pm 1.1 \right) \times 10^{-6} $ & \\
  &CLEO~~~~ $ \left (11.5_{-3.7-1.7}^{+4.5+1.8} \right) \times 10^{-6} $ & 
$ \left (10.6\pm 1.3 \right )\times 10^{-6}$ \\
    &Belle~~~~ $ \left( 10_{-1.5-0.8}^{+1.6+0.7} \right) \times 10^{-6} $ & \\\hline 
\end{tabular}
\caption{Experimental data of $B \to \phi K^\ast$ decays from BaBar 
\cite{babar_collab}, CLEO \cite{cleo_collab} and Belle \cite{belle_collab}
and their weighted average. }
\end{center}
\end{table}

The decay rate can be written as 
\beq
\Gamma(B \to V_1 V_2) = \frac{p_{cm}}{8 \pi m^2_B}
\left [\mid H_{00}\mid^2 + \mid H_{++}\mid^2 + \mid H_{--}\mid^2\right ] 
\eqn
Neglecting the annihilation contributions (which are expected to be small)
to $B \to \phi K^\ast $,  
$H_{00}$ and $H_{\pm\pm}$ are given by:
\beq
H_{00} &=& \frac{G_F}{\sqrt{2}}\frac{a^n(\phi K^\ast) f_\phi}{2 m_{K^\ast}}
\Bigg \{\left(m^2_B - m^2_{K^\ast}-m^2_\phi\right)\left(m_B+ m_{K^\ast}\right)
A^{BK^\ast}_1(m^2_\phi)\non \\
&& - \frac{4 m^2_B p^2_c}{m_B+m_{K^\ast}}
A^{BK^\ast}_2(m^2_\phi)\Bigg \}\non \\
H_{\pm\pm} &=& \frac{G_F}{\sqrt{2}}
a^n(\phi K^\ast)m_\phi f_\phi\Bigg \{\left(m_B+m_{K^\ast}\right)
A^{BK^\ast}_1(m^2_\phi) \non \\
&& \mp \frac{2m_B p_c}{m_B+m_{K^\ast}}
V^{BK^\ast}(m^2_\phi)\Bigg \}
\eqn
where, $a^n(\phi K^\ast)
= a^n_3 + a^n_4 + a^n_5 - (a^n_7 + a^n_9 + a^n_{10})/2 $.
The effective parameters $a_i$ appearing in the helicity amplitudes $H_{00}$
and $H_{\pm \pm}$ given in Ref.\cite{cheng-yang1, self} and other input 
parameters are given in Ref.\cite{self}.

In Table 1, we display the experimentally (BaBar, CLEO and Belle) 
measured branching ratios and the weighted averaged values 
for the $B^+ \to \phi K^{\ast +}$ and $\bar B^0 \to \phi \bar K^{\ast 0}$. The
theoretical predictions in the SM for two different form factor models, the
LCSR and BSW models are given in Ref.\cite{cheng-yang1}.  

\begin{table}
\begin{center}
\begin{tabular}{|c|c|c|c|}
\hline
$x$ & $(\rho, \psi )$ & ${\cal B}^{\rm SUSY}$ (in units of $10^{-6}$) 
& ${\cal A}_{CP}$ (in $\% $)  \\
\hline
  & ($0.4\times 10^{-2}$, -0.5) & $23.37^{+4.88}_{-4.42}~~(21.76^{+4.56}_{-4.13})$ 
& $-4.7 ( -4.4)$  \\
\cline{2-4}
0.3  & ($0.4\times 10^{-2}$, -0.7) & $21.50^{+4.49}_{-4.06}~~(20.17^{+4.22}_{-3.83})$ 
& $-7.0~~(-6.5) $  \\
\cline{2-4}
  & ($0.6\times 10^{-2}$, 1.5) & $27.46^{+5.75}_{-5.2}~~(26.82^{+5.62}_{-5.08}) $ 
& $17.7~~(15.7)$  \\
\hline
 & $ (2.4\times 10^{-2}, -0.5 ) $ & $ 24.33^{+5.08}_{-4.6}~~(22.65^{+4.75}_{-4.3})$ 
& $ -4.7~~(-4.4) $ \\
\cline{2-4}
4.0 & ($2.6\times 10^{-2}$, -0.45) & $ 26.98^{+5.6}_{-5.1}~~(25.11^{+5.27}_{-4.76})$ 
& $-4.2~~(-3.9) $ \\
\cline{2-4}
 & ($2.8\times 10^{-2}$, -0.8) & $ 25.31^{+5.3}_{-4.8}~~(23.87^{+5.01}_{-4.52})$ 
&$-8.0~~(-7.5)$  \\
\hline
\end{tabular}
\caption{${\cal B}(B^+ \to \phi K^{\ast +})$  and 
${\cal A}_{CP}(B^+\to \phi K^{\ast +})$ at the QCD scale 
$\mu = m_b$ for $LR$ mass insertion for selected points in the 
allowed $\rho-\psi $ space. The numbers in the parenthesis correspond to the
$RL$ mass insertion.   
The standard model branching ratio corresponding to this scale is 
$ (6.18^{+1.29}_{-1.15} ) \times 10^{-6}$. The errors are due to 
$\pm 10\%$ theoretical uncertainties in the calculation.}
\end{center}
\end{table}
\subsection{$LR(RL)$ mixing contributions to $B \to \phi K^\ast $ }

In this section we will study the effect of 
$LR(RL)$ mixing in $B \to \phi K^\ast $ process. This $LR(RL)$ mass insertion
can enhance the Wilson coefficient $C_{8g}$ by a factor of $m_{\tilde g}/m_b$
compared to the standard model contribution in the same way as shown in
section 5 for $B \to \phi K $ process. Hence, one need to impose the 
constrain on $LR(RL)$ mixing from experimentally measured 
${\cal B}(B \to X_s\gamma)$ and also from $S_{\phi K}, C_{\phi K}$ and 
${\cal B}(B \to \phi K)$ as obtained section 5. 

In this scenario, the new weak phase $\psi $, (the phase of the $LR(RL)$ 
mixing ) will contribute to direct CP-violating asymmetry
${\cal A}_{CP}$ defined as :
\beq
{\cal A}_{CP} = \frac{\Gamma (B^+ \to \phi K^{\ast +}) 
- \Gamma (B^- \to \phi K^{\ast -})} {\Gamma (B^+ \to \phi K^{\ast +}) 
+ \Gamma (B^- \to \phi K^{\ast -})} 
\eqn
in terms of partial widths. 
Recently BaBar and Belle Collaboration has presented their measurement of 
CP violating asymmetries for $B^0 \to \phi K^{\ast 0}$ and 
$B^\pm \to \phi K^{\ast \pm}$~\cite{babar_collab, belle_collab}
\beq
{\cal A}_{CP}(\bar B^0 \to \phi \bar K^{\ast 0} ) &=& 0.04\pm 0.12\pm0.02,~~~
0.07\pm0.15^{+0.05}_{-0.03}\\
{\cal A}_{CP}(B^\pm \to \phi K^{\ast \pm} ) &=&  +0.16\pm 0.17\pm 0.04,~~~
-0.13\pm 0.29^{+0.08}_{-0.11}
\eqn
where, in each asymmetry result, the first number correspond to the BaBar 
data while the second one correspond to Belle measurement. The standard
model value for this asymmetry is less than $1 \%$.
The new physics (SUSY) contributions from the new penguin 
operator appeared due to $LR(RL)$ mixing $(\delta^{bs}_{LR(RL)})$ can modify 
the sign and magnitude of ${\cal A}_{CP}(B^\pm \to \phi K^{\ast \pm})$ within 
the allowed parameter space of $\delta^{bs}_{LR(RL)}$. 


\begin{table}
\begin{center}
\begin{tabular}{|c|c|c|c|}
\hline
$x$ & $(\rho, \psi )$ & ${\cal B}^{\rm SUSY}$ (in units of $10^{-6}$) 
& ${\cal A}_{CP}$ (in $\% $)  \\
\hline
  & ($0.55\times 10^{-2}$, 1.8) & $31.83^{+6.6}_{-6.0}~~(32.45^{+6.7}_{-6.1})$ &
  $19.39^{+0.01}_{-0.02}~~(16.17^{+0.05}_{-0.07}) $ \\
\cline{2-4}
0.3 &($0.82\times 10^{-2}$, 2.8) & $ 14.50^{+3.04}_{-2.74}~~(21.62^{+4.44}_{-4.02}) $ &
$20.2~~(10.96^{+0.06}_{-0.08})$ \\
\cline{2-4}
   &($0.82\times 10^{-2}$, 2.9) & $ 12.39^{+2.59}_{-2.34}~~(19.82^{+4.05}_{-3.67})$ &
   $15.73~~(8.04^{+0.05}_{-0.06}) $ \\
\hline
\end{tabular}
\caption{ 
${\cal B}(B^+ \to \phi K^{\ast +})$
and ${\cal A}_{CP}(B^+\to \phi K^{\ast +})$ at the QCD scale 
$\mu = m_b/2$ for $LR$
mass insertion for selected points in the allowed $\rho-\psi $ space.
The numbers in the parenthesis correspond to the $RL$ mass insertion.  
The standard model branching ratio corresponding to this scale is 
$ (14.92^{+3.08}_{-2.78}) \times 10^{-6}$. The errors are due to 
$\pm 10\%$ theoretical uncertainties in the calculation.}
\end{center}
\end{table}
To get the numerical values of 
${\cal B}(B^+ \to \phi K^{\ast +})$, and
${\cal A}_{CP}( B^+\to \phi K^{\ast +}) $, we fix $x= 0.3 $ and $4.0$. Then
for a given QCD scale $\mu $, we select some points in the allowed parameter
space of $\rho-\psi$ plane (as marked by $Z$ in Figure~\ref{fig1}) for 
both values of $x$. In this computation, we 
include $(\pm 10\%)$ theoretical uncertainties. 

In Table~2, we present the branching ratio 
${\cal B}(B^+ \to \phi K^{\ast +})$ and the CP rate asymmetry 
${\cal A} (B^+\to \phi K^{\ast +}) $ 
for $\mu = m_b$ and selected values of $x = 0.3 $ and $4.0$ for 
values of $\rho $ and $\psi $ allowed by $B \to \phi K_S$ data 
for $LR$ mass insertion ($RL$ is shown in the
parenthesis). The branching ratio with SUSY 
turn out to be much higher than the standard model value 
of $6.18^{+1.29}_{-1.15}$, which is lower than the experimental 
data (Table~1). Even the lower range of theory 
prediction is much higher than the upper range of experimental data within
$1\sigma $. The rate asymmetry has much less error and is consistently within
the range $\sim -4\%$ to $\sim 18 \%$. 

Similarly in Table 3, we show ${\cal B}$ and ${\cal A}_{CP}$ calculated 
for QCD scale $\mu = m_b/2$. In this case, there are two allowed regions
from the combined $B \to \phi K $ and $B \to X_s \gamma$ constraints 
corresponding to $x = 0.3$. For $x = 4.0$, there are no allowed regions 
from $B \to \phi K $ data. The standard model branching ratio is much 
larger compared to the one computed at $\mu = m_b$. In SUSY, apart from 
the QCD scale $\mu$, the branching
ratio also depend on the values of $\rho $ and $\psi$. Moreover, the selected
points in $\rho-\psi$ plane are different in the two cases. In the case, 
with $\mu = m_b/2$, and at $\rho = 0.82\times 10^{-2}$ and $\psi = 2.8, 2.9$ radians, 
with $LR$ mixing, lower ranges of the theory predictions are consistent 
with the upper range of experimental 
data at one sigma. For the other value of $\rho$ and $\psi$, the theoretical 
prediction for branching ratio is much higher than the standard model theory
as well as experimental data. With $RL$ mixing, the predicted branching ratio
is much larger compared to both the standard model prediction and experimental
data. The asymmetries for both $LR$ and $RL$ mixing case are always positive 
with less errors.    

We conclude that for some selected points in $\rho-\psi$ plane allowed by 
$B \to \phi K $ and $B \to X_s \gamma$ at $\mu = m_b/2$ provide a satisfactory 
understanding of $B\to \phi K^\ast $ process. We also note that, at 
$\mu = m_b$ the SUSY contribution to the branching ratio of $B \to \phi K^{*}$ 
is too large to be consistent with the experimental data. 

We have also studied other CP violating asymmetries that can arise in 
vector-vector final state. The set of observables are defined 
in terms of $A_0, A_{\mid\mid}$ and $A_{\perp}$ as follows \cite{sinha-london}.
\beq
\Lambda_{\lambda} \, = \, \frac{| A_{\lambda}|^2
+ | \bar{A}_{\lambda}|^2}{2}, ~~~~\Sigma_{\lambda\lambda}
= \frac{\mid A_{\lambda}\mid^2 - \mid \bar{A}_{\lambda}\mid^2}{2},\non\\
 \Lambda_{\perp i} \, = \, - \,
{\rm Im} \left( A_{\perp} A_i^* - \bar{A}_{\perp}
\bar{A}_i^* \right) ,~~~~\Lambda_{\mid\mid 0}
= {\rm Re} \left ( A_{\mid\mid}A^{\ast}_{0} + \bar{A}_{\mid\mid} \bar{A^{\ast}}_{0}
\right ), \non \\
\Sigma_{\perp i} = -{\rm Im } \left (A_{\perp}A^{\ast}_{i}
+\bar{A}_{\perp}\bar{A^{\ast}}_{i}\right),
~~~~\Sigma_{\mid\mid 0} = {\rm Re} \left(A_{\mid\mid}A^{\ast}_{0}
-\bar{A}_{\mid\mid} \bar{A^{\ast}}_{0}\right ),\non\\
\rho_{\perp i} = {\rm Re} \left(\frac{q}{p}\left[A^{\ast}_{\perp}\bar{A}_{i}
+ A^{\ast}_{i}\bar {A}_{\perp}\right]\right ),
~~~~\rho_{\perp\perp} = {\rm Im}\left (\frac{q}{p} A^{\ast}_{\perp}
\bar{A}_{\perp}\right ), \non\\
\rho_{\mid\mid 0} = -{\rm Im}\left (\frac{q}{p}\left[A^{\ast}_{\mid\mid}
\bar{A}_{0}
+ A^{\ast}_{0}\bar {A}_{\mid\mid}\right]\right ),
~~~~\rho_{ii} = -{\rm Im}\left ( \frac{q}{p} A^{\ast}\bar{A}_{i}\right )
\eqn

where $\lambda = \lbrace 0, \, \parallel , \,
\perp \rbrace$ and the observables
where $i = \lbrace 0, \, \parallel \rbrace$, 
We restrict ourselves to the study of helicity dependent $CP$ asymmetry 
defined as $\Sigma_{\lambda\lambda}/\Lambda_{\lambda\lambda}$
~\cite{sinha-london}. For the purpose of illustration we select last 
two sample points from the Table 3. At these values of $\rho $ and $\psi$, 
with $LR$ mass insertion, the lower range of 
the ${\cal B}^{SUSY}$ is consistent with the upper range of the 
experimental data on ${\cal B}(B \to \phi K^{*})$ at one sigma. 
We then compute $\Sigma_{\lambda\lambda}/\Lambda_{\lambda\lambda}$ for 
each values of $\lambda $ for these two sets of $\rho $ and $\psi$ and 
is shown in Table 4. As before,  in this case also we include 
$\pm 10\%$ theoretical uncertainties in our calculation. We only show 
the helicity dependent asymmetries for $LR$ mass insertion, since with $RL$
mass insertion the SUSY contribution to the branching ratio is too large 
to be consistent with the data. 
\begin{table}
\begin{center}
\begin{tabular}{|c|c|c|c|c|}
\hline
$x$ & $(\rho, \psi )$ & $\Sigma_{00}/\Lambda_{00}$ 
& $\Sigma_{\mid\mid}/\Lambda_{\mid\mid}$ & $  
\Sigma_{\perp\perp}/\Lambda_{\perp\perp}$\\
\hline
0.3  & ($0.82\times 10^{-2}$, 2.8) & $0.19\pm 0.00$ & $ 0.61^{+0.009}_{-0.012}$ & 
$0.57^{+.011}_{-0.013}$ \\
\cline{2-5}
 & ($0.82\times 10^{-2}$, 2.9)& $0.15\pm 0.00$ & $ 0.71^{+0.020}_{-0.026}$ 
& $0.65^{+0.022}_{-0.027}$\\
\hline
\end{tabular}
\caption{Helicity dependent CP asymmetry at the QCD scale $\mu = m_b/2$ 
for $LR$ mass insertion for selected points in the allowed $\rho-\psi $ space.
The errors consist of $\pm 10\%$ theoretical uncertainties.}
\end{center}
\end{table}
\section{Conclusions}
In this talk, we considered the SUSY contribution to the gluonic dipole
moment operator to $ B\to \phi K_S$ process. We found that the $LR(RL)$
mass insertion can enhance the gluonic dipole moment operator significantly.
We then used the experimentally measured quantities, 
such as $S_{\phi K}, C_{\phi K}$ and ${\cal B}(B \to \phi K_S)$ to constrain
the parameter space of $LR(RL)$ mixing. Interestingly, we find that the 
constraints from $B \to \phi K$ data is consistent with the $B \to X_s\gamma$ 
limit. It turned out that the same enhancement of gluonic dipole
moment operator could also affect other penguin dominated process, such as 
$B \to \phi K^\ast $, which is a pure penguin process like 
$B \to \phi K_S$. In standard model, the predicted 
${\cal A}_{CP}(B \to \phi K^\ast )$ is less than $ 1 \%$. 
We calculated such asymmetries and also the branching ratio for the set of 
parameters allowed by $B \to \phi K$ data. 
At $\mu = m_b$, for both $LR$ and $RL$ mass insertion, we observed that  
the predicted branching ratio is well above the experimentally measured 
one. On the other hand, at the QCD scale $\mu = m_b/2$ 
with $LR$ mass insertion, we found that the theoretically 
computed branching ratio is consistent with the data with in one sigma error. 
At this second choice of allowed parameter space of $\delta^{bs}_{LR(RL)}$, 
we found $ {\cal A}_{CP}(B^+ \to \phi K^{\ast +})$ in the range 
$15\%$ to $20\%$, which is significantly higher than the standard model 
prediction but is still consistent with the present data. Finally, we also 
presented helicity dependent CP asymmetries in the same parameter space of 
$\delta^{bs}_{LR}$.
\begin{flushleft}
\begin{large}
{\bf Acknowledgments}
\end{large}
\end{flushleft}
This work was supported in part by US DOE contract numbers DE-FG03-96ER40969.

\def\pr#1,#2 #3 { {Phys.~Rev.}        ~{\bf #1},  #2 (19#3) }
\def\prd#1,#2 #3{ { Phys.~Rev.}       ~{D \bf #1}, #2 (19#3) }
\def\pprd#1,#2 #3{ { Phys.~Rev.}      ~{D \bf #1}, #2 (20#3) }
\def\prl#1,#2 #3{ { Phys.~Rev.~Lett.}  ~{\bf #1},  #2 (19#3) }
\def\pprl#1,#2 #3{ {Phys. Rev. Lett.}   {\bf #1},  #2 (20#3)}
\def\plb#1,#2 #3{ { Phys.~Lett.}       ~{\bf B#1}, #2 (19#3) }
\def\pplb#1,#2 #3{ {Phys. Lett.}        {\bf B#1}, #2 (20#3)}
\def\npb#1,#2 #3{ { Nucl.~Phys.}       ~{\bf B#1}, #2 (19#3) }
\def\pnpb#1,#2 #3{ {Nucl. Phys.}        {\bf B#1}, #2 (20#3)}
\def\prp#1,#2 #3{ { Phys.~Rep.}       ~{\bf #1},  #2 (19#3) }
\def\zpc#1,#2 #3{ { Z.~Phys.}          ~{\bf C#1}, #2 (19#3) }
\def\epj#1,#2 #3{ { Eur.~Phys.~J.}     ~{\bf C#1}, #2 (19#3) }
\def\eepj#1,#2 #3{ { Eur.~Phys.~J.}     ~{\bf C#1},#2 (20#3) }
\def\mpl#1,#2 #3{ { Mod.~Phys.~Lett.}  ~{\bf A#1}, #2 (19#3) }
\def\ijmp#1,#2 #3{{ Int.~J.~Mod.~Phys.}~{\bf A#1}, #2 (19#3) }
\def\ptp#1,#2 #3{ { Prog.~Theor.~Phys.}~{\bf #1},  #2 (19#3) }
\def\jhep#1, #2 #3{ {J. High Energy Phys.} {\bf #1}, #2 (19#3)}
\def\pjhep#1, #2 #3{ {J. High Energy Phys.} {\bf #1}, #2 (20#3)}


\end{document}